\documentclass[11pt,fleqn]{article}
\usepackage{graphicx}
\usepackage[a4paper,left=2cm,right=2cm]{geometry}
\usepackage{amsmath}
\usepackage{float}
\usepackage[latin1]{inputenc}
\usepackage{amsmath}
\usepackage{amsfonts}
\usepackage{amssymb}
\usepackage{float}
\floatstyle{boxed} 
\restylefloat{figure}

\newcommand{\be}{\begin{equation}}
\newcommand{\ee}{\end{equation}}
\newcommand{\bea}{\begin{eqnarray}}
\newcommand{\eea}{\end{eqnarray}}

\begin{document}
\title{Explaining the observed deviation in R($D^{(*)}$) in an anomalous 2HDM.}
\author{Lobsang Dhargyal \\\\\ Institute of Mathematical Sciences, Chennai 600113
, India.}
%\date{18 May 2017}

\maketitle
\begin{abstract}

The reported combined excess of about 4$\sigma$, by BABAR, Belle and LHCb, in the measurements of R($D^{(*)}$) and Br(B $\rightarrow \tau \nu_{\tau}$) from that of the standard model expectation constitute the most significant discrepancy from the standard model in collider experiments thus far(apart from the non zero neutrino masses). Here we analyze the phenomenological implications for these decay modes in a Flipped/Lepton-Specific two-Higg-doublet model with anomalously enhanced charged Higgs coupling to $\tau$/b. We conclude that this extension of the standard model can give results in agreement within 1$\sigma$ of the experimental values for the combination of R($D^{(*)}$) and Br(B $\rightarrow \tau \nu_{\tau}$) data.

\end{abstract}

\section{\large Introduction.}

It has been reported first by Babar and Belle, a possible deviation from the lepton flavor universality in $R(D^{(*)}) = \frac{Br(B \rightarrow D^{(*)}\tau\nu)}{Br(B \rightarrow D^{(*)}l\nu)}$. The present world average gives a deviation of 1.8$\sigma$ for the R(D) and 3.3$\sigma$ for the $R(D^{*})$ and the taken together corresponds close to a 3.8$\sigma$ deviation from SM prediction. There is also deviation reported in Br($B \rightarrow \tau \nu$) which is 1.3$\sigma$ above the SM prediction. When the errors in $R(D^{*})$ and $B(B \rightarrow \tau \nu)$ are added in quadrature, deviation from SM is about 4$\sigma$. In this work we present a Higgs mediated flavor universality violation in a Flipped or Lepto-Specific 2HDM with anomalous charged Higgs coupling to $\tau$ lepton or b quark respectively. This work is base on the materials contain in the reference \cite{our}.

\section{Results}

Due to limited space given, we will only show the main results from \cite{our}. In Table-\ref{Tab:one} we have shown for two different values of the parameters from the fits to the data :\\
\\
\begin{table}[H]
\begin{center}
\begin{tabular}[b]{|c|c|c|c|c|c|} \hline
S.no & $\tan\beta$ & $M_{\pm}$ GeV & $R(D)_{Th}$ & $R(D^{*})_{Th}$ & $Br_{Th}(B \rightarrow \tau\nu)$ \\
\hline\hline
1 & 69.97 & 700 & 0.348 & 0.255 & 1.29$\times10^{-4}$ \\
\hline
2 & 99.95 & 1000 & 0.348 & 0.255 & 1.29$\times10^{-4}$ \\
\hline
\end{tabular}
\end{center}
\caption{$\chi^{2}_{min} = 10.95$ and we have taken the $\tan\beta$ as $100 > \tan\beta > 1$.}
\label{Tab:one}
\end{table}

\section{Conclusions}

By adding theoretical and experimental errors in quadrature \cite{our}, we conclude that our model agrees within 1$\sigma$ for the combination of $R(D^{(*)})$ and $\mathcal{B}r(B \rightarrow \tau \nu_{\tau})$ data compare to about 4$\sigma$ deviation from SM predictions. The same results can be achieved if b quark replaces the $\tau$ lepton in a Lepton Specific 2HDM. From the form of the Yukawa couplings it is expected that if we require $\eta = -1$ for the b quark or $\tau$ lepton in the 2HDM-II will also work, an anomalous SUSY?

\section*{Acknowledgements}

Author would like to thank Nita Sinha and Rahul Srivastava, Institute of Mathematical Sciences for helpful discussions and comments. Author would also like to thank Shrihari Gopalakrishna, Institute of Mathematical Sciences for helpful comments. This work is supported and funded by the Department of Atomic Energy of the Government of India and by the Government of Tamil Nadu.

\end{document}